\documentclass[twocolumn,prc,aps,floatfix]{revtex4}
\usepackage[dvips]{graphicx}
\begin{document}
\newcommand{\be}{\begin{eqnarray}}
\newcommand{\ee}{\end{eqnarray}}
\title{Enhanced effects of the Lorentz invariance and  Einstein's equivalence principle violation  in 7.8 eV $^{229}$Th nuclear transition}
\author{V.V. Flambaum}
\affiliation{
 School of Physics, The University of New South Wales, Sydney NSW
2052, Australia
}
\date{\today}
\begin{abstract}
The Lorentz invariance and Einstein equivalence principle violating effects  in the narrow 7.8 eV transition in  $^{229}$Th nucleus may be  $10^5$ times larger than in atoms. This transition may be investigated using high precision laser spectroscopy methods, has a very small width, and suppressed  systematic effects. Similar values of the effects  are expected in 73 eV $^{235}$U nuclear transition which is coming within the reach of the laser spectroscopy.  M\"ossbauer transitions give another possibility.
\end{abstract}
\maketitle
%\pacs{PACS: 06.20.Jr, 42.62.Fi, 23.20.-g}
PACS numbers: 11.30.Er,  12.60.-i,  42.62.Fi , 23.20.Lv
%11.30.Er 	Charge conjugation, parity, time reversal, and other discrete symmetries
%12.60.-i	Models beyond the standard model
%42.62.Fi	Laser spectroscopy
%23.20.Lv	? transitions and level energies
%\section{Introduction}
Measurements of  the  Local Lorentz Invariance Violation (LLIV) effects is a very popular direction of the search for physics beyond the Standard Model (see, e.g. ~\cite{KosRus11,KosteleckyPottingPRD1995,Horava,Pospelov}). The LLIV effects are usually classified using notations of the works \cite{KosRus11,KosLan99,Colladay1997,Colladay1998} coming under the name the Standard Model Extension (SME). In the present paper we are interested in the term violating Lorentz invariance and Einstein's equivalence principle which in the non-relativistic limit is  presented as 
\begin{equation}\label{eq1}
\delta H=-\left(  C_{0}^{(0)}-\frac{2U}{3c^{2}}c_{00}\right)
\frac{\mathbf{p}^{2}}{2m},
%-\frac{1}{6}C_{0}^{(2)}T^{(2)}_{0},\label{eq1}
\end{equation}
where $\mathbf{p}$ is the momentum, $c$ is the speed of light, and $U$ is
the Newtonian gravitational potential. To measure this term we should consider transitions where the kinetic energy $\mathbf{p}^2/2m$ changes. 

 A high precision laser atomic spectroscopy have been used  to establish limits  on the interaction (\ref{eq1} )\cite{Pruttivarasin2014,Hohensee2013c,Dzuba2015}.  Using the virial theorem  for the Coulomb interaction we obtain the difference of kinetice energies between  excited and ground states\\  
$-(<\mathbf{p}^2/2m_e>_{exc} - <\mathbf{p}^2/2m_e>_{gr})=  \hbar \omega \sim$ few eV. 

Note, that the relativistic corrections to the electron wave functions and energies such as that produced by the spin-orbit interaction, are quite large  in heavy atoms ($\sim Z^2 \alpha^2$) and  change this estimate dramatically for  small $\hbar \omega$. For example, the difference $-(<\mathbf{p}^2/2m_e>_{exc} - <\mathbf{p}^2/2m_e>_{gr})$ does not tend to zero for $\hbar \omega=0$ \cite{Hohensee2013c}. However, the relativistic corrections to the LLIV operator Eq. (\ref{eq1}) remain very small \cite{Hohensee2013c,Dzuba2015}.

To have $10^5$ larger difference\\  $<\mathbf{p}^2/2m>_{exc} - <\mathbf{p}^2/2m>_{gr}$ and higher sensitivity to LLIV we propose to use a narrow ultraviolet   transition between the ground state and first metastable excited state in  $^{229}$Th nucleus. Currently this transition is very intensely investigated  both theoretically and experimentally. A brief review with  numerous references may be found in the recent paper \cite{ Tkalya2015}. The latest transition energy measurement gave $\omega =$ 7.8(5) eV \cite{Beck}.  The estimated width is $\sim 10^{-3}-10^{-4}$ Hz \cite{ Tkalya2015,th2}. 
In Refs. \cite{th4,Campbell,Rellergert,Peik2015} it was suggested to use this transition as a high precision nuclear clock with a projected accuracy exceeding that of the current atomic clocks by two orders of magnitude \cite{Campbell}. This transition may also be used to build  a nuclear laser  \cite{Tkalya-11}.  Possible effects of the space-time variation  of the fundamental constants such as the fine structure constants $\alpha$, quark masses and the strong interaction scale $\Lambda_{QCD}$, are enhanced in this transition up to five orders of magnitude \cite{th1} (see also  \cite{He,Hayes,Wiringa,Litvinova,Dmitriev}).

  The ground state of  $^{229}$Th nucleus is $J^P[Nn_z\Lambda]=5/2^+[633]$;
i.e. the deformed oscillator quantum numbers are $N=6$, $n_z=3$, the
 projection of the valence neutron orbital angular momentum on the
nuclear symmetry axis (internal  z-axis)
 is $\Lambda=3$, the spin projection $\Sigma=-1/2$, and the
 total angular momentum and the total 
 angular momentum projection are $J=\Omega=\Lambda+\Sigma=5/2$.
The 7.8 eV excited state is $J^P[Nn_z\Lambda]=3/2^+[631]$; i.e.
it has 
$\Lambda=1$, $\Sigma=1/2$ and  $J=\Omega=3/2$.

The energy of both states
may be described by an equation \\
$E=<\frac{p^2}{2m_n}> +<U> + C \Lambda \Sigma$,\\
where $m_n$ is the neutron mass and  the spin-orbit interaction constant in Th nucleus is $C= - 0.85$ MeV \cite{BM}.
In the simplest single-particle model  the difference of the spin-orbit energies between the excited ( $\Lambda=1$, $\Sigma=1/2$) and the ground ($\Lambda=3$, $\Sigma=-1/2$) states is $2 C$.
The many-body corrections reduce this difference to $1.2 C$ \cite{Wiringa}.

In the oscillator potential the kinetic and  potential energies are equal, $<\frac{p^2}{2m}> =<U>$.
This gives us an estimate for the  the difference of kinetic energies between  the excited and ground states\\  
$<\mathbf{p}^2/2m_n>_{exc} - <\mathbf{p}^2/2m_n>_{gr}=  0.5 (\hbar \omega -1.2 C)= 0.5$ MeV.

This simple analytical estimate shows that the Lorentz invariance and Einstein equivalence principle violating effects Eq.(\ref{eq1}) in $^{229}$Th nucleus may be  $10^5$ times larger than in atoms. The 7.8 eV transition may be investigated using high precision laser spectroscopy methods, has a very small width, extremely small black body radiation shift and suppressed other systematic effects  (see e.g. \cite{Campbell}).
This gives additional advantages. We conclude
that this nuclear transition has a great potential for a laboratory search for new physics. 
 
  A similar value 
  $<\mathbf{p}^2/2m_n>_{exc} - <\mathbf{p}^2/2m_n>_{gr} \sim $ 0.1 - 1 MeV
  is expected in $^{235}$U 73 eV transition. The laser spectroscopy  for this transition will be available soon \cite{Ye}.  Another possibility is M\"ossbauer transitions. However, the frequencies in these transitions are larger, so the relative effects will be smaller. The accuracy of the measurements in the M\"ossbauer spectroscopy is also not as high as in the laser spectroscopy

This work is  supported by the Australian Research Council.


\begin{thebibliography}{99}
 \bibitem{KosRus11}Kosteleck{\'y}, V.~A. \& Russell, N.
\newblock{Data tables for Lorentz and CPT violation.}
\newblock {\em Rev. Mod. Phys.} \textbf{83}, 11-32 (2011). (Updated yearly: \newblock{\em
arXiv:0801.0287v8}.)

\bibitem{KosteleckyPottingPRD1995}
Kosteleck\'{y}, V. A. \& Potting, R.
\newblock{CPT, strings, and meson factories.}
\newblock{\em Phys. Rev. D} \textbf{51}, 3923-3935 (1995).

\bibitem{Horava} Horava, P.
\newblock{Quantum gravity at a Lifshitz point.}
\newblock{\em Phys. Rev. D} \textbf{79}, 084008 (2009).

\bibitem{Pospelov} Pospelov, M. \& Shang, Y.
\newblock{Lorentz violation in Horava-Lifshitz-type theories.}
\newblock{\em Phys. Rev. D} \textbf{85}, 105001 (2012).

\bibitem{KosLan99} Kosteleck{\'y}, V.~A. \& {Lane}, C.~D.
\newblock{Constraints on Lorentz violation from clock-comparison experiments.}
\newblock{\em Phys. Rev. D} \textbf{60}, 116010 (1999).


\bibitem{Colladay1997} Colladay, D. \& Kosteleck\'y, V. A.
\newblock{CPT violation and the standard model.}
\newblock{\em Phys. Rev. D} \textbf{55}, 6760-6774 (1997).

\bibitem{Colladay1998} Colladay, D. \& Kosteleck\'y, V. A.
\newblock{Lorentz-violating extension of the standard model.}
\newblock{\em Phys. Rev. D} \textbf{58}, 116002 (1998).


\bibitem{Hohensee2013c}
Hohensee, M. A. \textit{et al.}
\newblock {Limits on Violations of Lorentz Symmetry and the Einstein Equivalence Principle
using Radio-Frequency Spectroscopy of Atomic Dysprosium}.
\newblock {\em Phys. Rev. Lett.} \textbf{111}, 050401 (2013).

\bibitem{Pruttivarasin2014} Pruttivarasin, T. \textit{et al.}
\newblock{Michelson-Morley analogue for electrons using trapped Ions to test Lorentz
symmetry.}
\newblock {\em Nature} 517, 592-595 (2015).


\bibitem{Dzuba2015} V. A. Dzuba, V. V. Flambaum, M. Safronova, S. G. Porsev, T. Pruttivarasin,  M. A. Hohensee, H. H\"affner,
%\textit{et al.}
Nature-Physics, 2015 (in press),  arxiv: 1507.06048

\bibitem{Tkalya2015} E. V. Tkalya,C. Schneider, J. Jeet, and E. R. Hudson, arxiv: 1509.09101 

\bibitem{Beck} 
B. R. Beck, J. A. Becker, P. Beiersdorfer, G. V. Brown, K. J. Moody, 
J. B. Wilhelmy, F. S. Porter, C. A. Kilbourne, and R. L. Kelley,
Phys. Rev. Lett. {\bf 98}, 142501 (2007). Update value 7.8(5) eV is presented at URL https://e-reports-ext.llnl.gov/pdf/375773.pdf

\bibitem{th2}
E. V. Tkalya, A. N. Zherikhin, and V. I. Zhudov, 
Phys. Rev. C {\bf 61}, 064308 (2000);
A. M. Dykhne, E. V. Tkalya, 
Pis'ma Zh. Eks. Teor. Fiz. {\bf 67}, 233 (1998)
[JETP Lett. {\bf 67}, 251 (1998)].

\bibitem{th4}
E. Peik and Chr. Tamm, Europhys. Lett. {\bf 61}, 181 (2003).

\bibitem{Campbell} C. J. Campbell, A. G. Radnaev, A. Kuzmich, V. A.
Dzuba, V. V. Flambaum, and A. Derevianko, Phys. Rev.
Lett. 108, 120802 (2012).

\bibitem{Rellergert} W. G. Rellergert, D. DeMille, R. R. Greco, M. P. Hehlen,
J. R. Torgerson, and E. R. Hudson, Phys. Rev. Lett. 104,
 200802 (2010).

\bibitem{Peik2015}  E. Peik and M. Okhapkin, C. R. Phys. 16, 516 (2015),

\bibitem{Tkalya-11}
E.~V. Tkalya,
  Phys. Rev. Lett. \textbf{106},
  162501, 2011.
  
\bibitem{th1}
V. V. Flambaum,  Phys. Rev. Lett. {\bf 97}, 092502 (2006).

\bibitem{He}
[9] H.-t. He and Z.-z. Ren, J. Phys. G: Nucl. Phys. {\bf 34}, 1611
(2007).
\bibitem{Hayes} A. C. Hayes and J. L. Friar, Phys. Lett. B {\bf 650}, 229
(2007).

\bibitem{Wiringa} V.V. Flambaum, R.B. Wiringa, 
 Phys. Rev. C {\bf 79}, 034302 (2009).

\bibitem{Litvinova} E. Litvinova, H. Feldmeier, J. Dobaczewski, and V. Flambaum,
Phys. Rev. C {\bf 79}, 064303 (2009).

\bibitem{Dmitriev} V.V. Flambaum, N. Auerbach, V.F. Dmitriev
 EPL (Europhysics Letters) {\bf 85}, 50005 (2009)

\bibitem{BM} A. Bohr, B. R. Mottelson. Nuclear structure, V. 2.
(Benjamin, New York, Amsterdam, 1974).\\
Note that the spin-orbit interaction in nuclei is significantly larger than that given by the Dirac equation (it has a different origin).

\bibitem{Ye}A. Cing\"oz,  D. C. Yost,  T. K.  Allison, A.  Ruehl, M.E.  Fermann, I.  Hartl, and  J. Ye, J. ,  Nature,  {\bf 482},    68 - 71  (2012).

\end{thebibliography}
\end{document}